%% file: sample-sigconf-authordraft.tex
\PassOptionsToPackage{table,xcdraw}{xcolor}
\documentclass[sigconf]{acmart}

\AtBeginDocument{%
  }

\copyrightyear{2026}
\acmYear{2026}
\setcopyright{cc}
\setcctype{by}
\acmConference[MOBILESoft '26]{IEEE/ACM 13th International Conference on Mobile Software Engineering and Systems}{April 12, 2026}{Rio de Janeiro, Brazil}
\acmBooktitle{IEEE/ACM 13th International Conference on Mobile Software Engineering and Systems (MOBILESoft '26), April 12, 2026, Rio de Janeiro, Brazil}
\acmDOI{10.1145/3795077.3795118}
\acmISBN{979-8-4007-2486-2/2026/04}

\usepackage[table,xcdraw]{xcolor}
\usepackage[framemethod=TikZ]{mdframed}
\usepackage{float}
\usepackage[utf8]{inputenc}
\usepackage[T1]{fontenc}
\usepackage{subcaption}
\usepackage{listings}
\usepackage{stackengine} 
\usepackage{tabularx}
\usepackage{diagbox}
\usepackage{adjustbox}
\definecolor{cyan1}{RGB}{0,255,255}
\definecolor{codegreen}{rgb}{0,0.6,0}
\definecolor{codegray}{rgb}{0.5,0.5,0.5}
\definecolor{codepurple}{rgb}{0.58,0,0.82}
\definecolor{backcolour}{rgb}{0.95,0.95,0.92}
\usepackage{hyperref}
\usepackage[nameinlink]{cleveref}
\usepackage{cleveref}
\usepackage{tikz} 
\usepackage{array,multirow}
\usepackage{rotating}
\usepackage{todonotes}
\definecolor{lightblue}{RGB}{200, 220, 255}
\definecolor{vsdarkbackground}{RGB}{30, 30, 30}
\definecolor{vsgreybackground}{RGB}{77,74,60}
\definecolor{vsdarkblue}{RGB}{94, 155, 255}
\definecolor{vsdarkgreen}{RGB}{120, 195, 80}
\definecolor{vsdarkred}{RGB}{200, 0, 0}
\definecolor{vsdarkpurple}{RGB}{177, 97, 211}
\definecolor{vsdarkgray}{RGB}{155, 155, 155}
\definecolor{lightblue}{RGB}{200, 220, 255}
\definecolor{lightgrey}{RGB}{230, 230, 230}
\definecolor{lightorange}{RGB}{255, 201, 102}
\definecolor{lightyellow}{RGB}{255, 255, 0}
\definecolor{loggray}{RGB}{160,160,160}
\definecolor{codegray}{rgb}{0.5,0.5,0.5}
\definecolor{codepurple}{rgb}{0.58,0,0.82}
\definecolor{backcolour}{rgb}{0.95,0.95,0.92}
\definecolor{lightred}{RGB}{255, 110, 110}
\definecolor{lightgreen}{RGB}{181, 230, 29}
\definecolor{codegreen}{rgb}{0,0.6,0}

\newcommand{\stilltodo}[1]{\textcolor{red}{#1}}

\newcommand{\directquote}[1]{\textit{``#1''}}

\makeatother

\lstdefinestyle{mystyle}{   
    commentstyle=\color{codegreen},
    keywordstyle=\color{magenta},
    numberstyle=\tiny\color{codegray},
    stringstyle=\color{codepurple},
    basicstyle=\ttfamily\footnotesize,
    breakatwhitespace=false,         
    breaklines=true,                 
    captionpos=b,   
    frame=single,
    keepspaces=true,                 
    numbers=left,                    
    numbersep=5pt,                  
    showspaces=false,                
    showstringspaces=false,
    showtabs=false,                  
    tabsize=2
}

\makeatother

\begin{document}

\title{Challenges in Android Data Disclosure: An Empirical Study} % and a Tool-Based Approach}

\author{Mugdha Khedkar}
\affiliation{%
  \institution{\textit{Heinz Nixdorf Institute \\ Paderborn University}}
  \city{Paderborn}
  \country{Germany}
}
\email{mugdha.khedkar@uni-paderborn.de}

\author{Michael Schlichtig}
\affiliation{%
  \institution{\textit{Heinz Nixdorf Institute \\ Paderborn University}}
  \city{Paderborn}
  \country{Germany}
}
\email{michael.schlichtig@uni-paderborn.de}

\author{Mohamed Soliman}
\affiliation{%
  \institution{\textit{Heinz Nixdorf Institute \\ Paderborn University}}
  \city{Paderborn}
  \country{Germany}
}
\email{mohamed.soliman@uni-paderborn.de}

\author{Eric Bodden}
\affiliation{%
  \institution{\textit{Heinz Nixdorf Institute \\ Paderborn University and Fraunhofer IEM}}
  \city{Paderborn}
  \country{Germany}
}
\email{eric.bodden@uni-paderborn.de}


\begin{abstract}
%\mk{I don't like this title because it is unclear how we are unifying app development with privacy assessments. How about ``Towards Automation of Privacy Assessments''?}

Current legal frameworks enforce that Android developers accurately report the data their apps collect. 
However, large codebases can make this reporting challenging. 
This paper employs an empirical approach to understand developers' experience with Google Play Store’s Data Safety Section (DSS) form. %, and explore the challenges they encounter when completing the form. 

We first survey 41 Android developers to understand how they categorize privacy-related data into DSS categories and how confident they feel when completing the DSS form. 
To gain a broader and more detailed view of the challenges developers encounter during the process, we complement the survey with an analysis of 172 online developer discussions, capturing the perspectives of 642 additional developers. 
Together, these two data sources represent insights from 683 developers.

%First, we survey 41 Android developers to capture first-hand experiences in completing the DSS form. 
%Next, we analyze 164 developer community discussions, incorporating the perspectives of 618 developers to reveal broader, community-level struggles with DSS reporting. 
%Together, these sources provide insights from a total of 659 developers.

Our findings reveal that developers often manually classify the privacy-related data their apps collect into the data categories defined by Google—or, in some cases, omit classification entirely—and rely heavily on existing online resources when completing the form. 
Moreover, developers are generally confident in recognizing the data their apps collect, yet they lack confidence  in translating this knowledge into DSS-compliant disclosures. 
Key challenges include issues in identifying privacy-relevant data to complete the form, limited understanding of the form, and concerns about app rejection due to discrepancies with Google’s privacy requirements. 
%We also identify several recurring challenges, 

These results underscore the need for clearer guidance and more accessible tooling to support developers in meeting privacy-aware reporting obligations.
\end{abstract}

\begin{CCSXML}
<ccs2012>
   <concept>
       <concept_id>10002978.10003022.10003027</concept_id>
       <concept_desc>Security and privacy~Social network security and privacy</concept_desc>
       <concept_significance>500</concept_significance>
       </concept>
   <concept>
       <concept_id>10011007.10011006.10011073</concept_id>
       <concept_desc>Software and its engineering~Software maintenance tools</concept_desc>
       <concept_significance>300</concept_significance>
       </concept>
 </ccs2012>
\end{CCSXML}

\ccsdesc[500]{Security and privacy~Social network security and privacy}
\ccsdesc[300]{Software and its engineering~Software maintenance tools}%%
\keywords{static analysis, data collection, data protection, privacy-aware reporting}


\maketitle

\input{Sections/Introduction}
\input{Sections/Background}
\input{Sections/Methodology}
\input{Sections/Results}

\input{Sections/Discussion}
\input{Sections/RelatedWork}
\input{Sections/Limitations}
\input{Sections/Conclusion}

\begin{acks}
This work is partially funded by the Deutsche Forschungsgemeinschaft (DFG, German Research Foundation), SFB 1119 (236615297). 
We thank Shashi Prakash for his help in conducting the survey, and Rahul Kanekar for his help with statistical analysis of the survey responses. 
We also thank the anonymous reviewers for their feedback and suggestions.
\end{acks}

\bibliographystyle{ACM-Reference-Format}
\bibliography{sample-base}

\begin{appendix}
\section{Codebook}
\label{codebook}
In this section, we present our codebook.

\begin{table}[h]
\small
\begin{tabular}{|c|c|l|}
\hline
\textbf{Theme}                                                                              & \textbf{Subtheme}                                                                                                                 & \multicolumn{1}{c|}{\textbf{Code(s)}}                                                                                           \\ \hline
\multirow{5}{*}{\begin{tabular}[c]{@{}c@{}}DSS \\support\end{tabular}}                                                                & \multirow{2}{*}{Online resources}                                                                                                 & Official documentation                                                                                                          \\ \cline{3-3} 
                                                                                            &                                                                                                                                   & Online forums                                                                                                                   \\ \cline{2-3} 
                                                                                            & \multirow{3}{*}{Tool}                                                                                                             & Checks framework                                                                                                                \\ \cline{3-3} 
                                                                                            &                                                                                                                                   & Third-party tool                                                                                                                \\ \cline{3-3} 
                                                                                            &                                                                                                                                   & In-house tool                                                                                                                   \\ \hline

\multirow{5}{*}{\begin{tabular}[c]{@{}c@{}}Confidence \\levels \\ of \\developers\end{tabular}} & \begin{tabular}[c]{@{}c@{}}Understanding collected data\end{tabular}                                                           & \multirow{5}{*}{\begin{tabular}[c]{@{}l@{}}Lacks confidence, \\ Uncertain, \\ Some confidence, \\ Fully confident\end{tabular}} \\ \cline{2-2}
                                                                                            & \begin{tabular}[c]{@{}c@{}}Understanding DSS \\definitions\end{tabular}                                                          &                                                                                                                                 \\ \cline{2-2}
                                                                                            & \begin{tabular}[c]{@{}c@{}}Classifying collected data\end{tabular}                                                             &                                                                                                                                 \\ \cline{2-2}
                                                                                            & \begin{tabular}[c]{@{}c@{}}DSS adhering to GDPR\end{tabular}                                                                   &                                                                                                                                 \\ \cline{2-2}
                                                                                            & \begin{tabular}[c]{@{}c@{}}Completing DSS form\end{tabular}                                                                    &                                                                                      \\ \hline                                          \multirow{14}{*}{\begin{tabular}[c]{@{}c@{}}DSS \\challenges\end{tabular}}                                                            & \multirow{4}{*}{\begin{tabular}[c]{@{}c@{}}Understanding \\ and complying \\ with DSS form\\\end{tabular}}                          & \begin{tabular}[c]{@{}l@{}}Confusing terminologies\end{tabular}                                                  \\ \cline{3-3} 
                                                                                            &                                                                                                                                   & Data deletion                                                                                                                   \\ \cline{3-3} 
                                                                                            &                                                                                                                                   & Encryption declaration                                                                                                          \\ \cline{3-3} 
                                                                                            &                                                                                                                                   & Legal compliance                                                                                                                \\ \cline{2-3} 
                                                                                            & \multirow{3}{*}{\begin{tabular}[c]{@{}c@{}}Identifying usage  of \\privacy-relevant  data \\to complete DSS form\end{tabular}} & Data collection                                                                                                                 \\ \cline{3-3} 
                                                                                            &                                                                                                                                   & TPL data collection                                                                                                             \\ \cline{3-3} 
                                                                                            &                                                                                                                                   & Data sharing                                                                                                                    \\ \cline{2-3} 
                                                                                            & \multirow{3}{*}{\begin{tabular}[c]{@{}c@{}}Technical process \\ support\end{tabular}}                                          & \begin{tabular}[c]{@{}l@{}}Technical submission \\process\end{tabular}                                                         \\ \cline{3-3} 
                                                                                            &                                                                                                                                   & Version control                                                                                                                 \\ \cline{3-3} 
                                                                                            &                                                                                                                                   & Lack of tool support                                                                                                            \\ \cline{2-3} 
                                                                                            & \begin{tabular}[c]{@{}c@{}}Understanding app rejection\end{tabular}                                                            & App rejection                                                                                                                   \\ \cline{2-3} 
                                                                                            & \multirow{3}{*}{\begin{tabular}[c]{@{}c@{}}App rejection \\ categories\end{tabular}}                                              & Missing information                                                                                                             \\ \cline{3-3} 
                                                                                            &                                                                                                                                   & \begin{tabular}[c]{@{}l@{}}Mismatch with \\ privacy policy\end{tabular}                                                         \\ \cline{3-3} 
                                                                                            &                                                                                                                                   & \begin{tabular}[c]{@{}l@{}}Mismatch with \\ app behavior\end{tabular}                                                           \\ \hline
\end{tabular}
\end{table}
\end{appendix}
\end{document}

%% file: Sections/Introduction.tex
\section{Introduction}
\label{intro}

Android apps are central to modern digital life, with over 1.5 million apps on the Google Play Store~\cite{apptrends} collecting sensitive data such as location, health, and financial information. 
While this data enables personalized services, it also raises privacy concerns. 

To address privacy concerns, the General Data Protection Regulation (GDPR)~\cite{gdpr}, enforced by the European Union on 25th May 2018, establishes strict rules on how personal data is collected, stored, and processed~\cite{art4}, with significant penalties for non-compliance~\cite{penalties}.  
Among its key provisions, Article §13~\cite{art13} mandates that users be informed of the collection and processing of personal data through documents such as privacy policies. 
However, such privacy policies are long and vague, and may not always be authored by the app developers. 
Several studies~\cite{privacypolicytrust,automatedriskanalysis,guileak,ppviolationappcode,PTPDroid} have consistently shown discrepancies between privacy policies and the actual source code, undermining their accuracy and misleading users. 

In response to these discrepancies, researchers proposed the use of privacy labels~\cite{2009label1,2009label2} to make data-handling practices more transparent to end users.  
Building on this idea, major app marketplaces such as Google Play Store~\cite{googleplaystore} and Apple Store~\cite{appstore} introduced mandatory data disclosure requirements. 
In 2022, Google launched the Data Safety Section (DSS)~\cite{data}, a privacy label that requires Android developers to declare how their apps collect, share, and protect user data. 
As part of this requirement, developers must complete a DSS form in the Google Play Console~\cite{playconsole}, disclosing information about data collection, sharing, and security practices across predefined data categories (e.g., location, personal information, financial information). 
%The form organizes information into broad data categories (e.g.,location, personal information, financial information), which are further divided into specific data types. 
%See \Cref{background} for more details. 
Since its introduction, Google has intensified enforcement of DSS compliance. 
In 2024 alone, the company unpublished over 1.3 million apps from the Play Store for excessive data collection—exceeding the number of new apps released that year~\cite{forbesstudy,apptrends}. 
Despite these regulatory and platform-level efforts, empirical studies~\cite{mozilla,datalabels,girish2025signaldataempiricalprivacy,khedkar2024androidappdevelopersaccurately,encryptionlabels} continue to find widespread inaccuracies in how data collection is reported through the DSS. 
For example, our previous study~\cite{khedkar2024androidappdevelopersaccurately,Khedkar2026} revealed that none of the 20 most downloaded apps on Google Play Store correctly disclosed their data collection practices in the DSS. 
While such findings highlight widespread misreporting, a clear \textbf{research gap} remains in understanding \emph{why} developers struggle with or fail to comply with DSS requirements.

To address this \textbf{research gap}, this paper performs an empirical study with the \textbf{goal} to \emph{explore how developers classify the privacy-related data their apps collect into DSS data categories, how confident they are when completing the DSS forms, and the challenges they encounter along the way}. 
By achieving this goal, we aim to identify opportunities to design improved tools, documentation, and processes that better support developers in producing accurate privacy disclosures.  %our study seeks to answer the following research questions: 

%\textbf{RQ1.} What methods and resources do developers use to categorize privacy-related data collected by their app into the DSS data categories? 

%\textbf{RQ2.} How confident are developers in their ability to accurately and effectively complete the DSS forms?

%\textbf{RQ3.} What are the challenges developers face when completing the DSS forms?

To achieve this goal, we design our study based on two data sources: \emph{developer survey} and \emph{online discussions}. 
We justify the choice of these data sources in \Cref{methodology_MMS}. 
First, we conduct an online survey targeting Android developers to gather in-depth insights with a total of \emph{\textbf{41} participants}.  
Next, we analyze 172 developer discussions across Stack Overflow, Reddit, Discord, GitHub, and Hacker News, allowing us to capture perspectives from Android developers who engage with peer communities to navigate DSS reporting with overall \emph{\textbf{642} unique developers}. 
Our research questions are defined in \Cref{methodology_MMS}.
%\eb{N = number of developers contributing to the discussions?}
%\mk{Yes}

In summary, this study makes the following contributions:
\begin{itemize}
	\item Our findings reveal that developers use manual \textbf{methods} to classify the privacy-related data their apps collect into DSS data categories—or, in some cases, omit classification entirely—and rely heavily on existing online \textbf{resources} when completing the DSS form. 
	\item We find that developers are generally \textbf{confident} in recognizing the data their apps collect, yet their confidence drops significantly when translating this knowledge into DSS-compliant disclosures. % \textbf{(RQ2)}.  
    \item We also identify several recurring \textbf{challenges}, including issues in identifying privacy-relevant data to complete the DSS form, limited understanding of the form, and concerns about app rejection due to discrepancies with Google’s privacy requirements. % \textbf{(RQ3)}.
\end{itemize}

All artifacts are available online~\cite{artifacts}. 
The rest of the paper is organized as follows: %Section~\ref{motivation} discusses a motivating example. 
\Cref{background} introduces some background about Google's Data Safety Section (DSS).  
In \Cref{methodology_MMS}, we explain the methodology of the empirical study, and discuss its results in \Cref{results_MMS}. 
%\Cref{autoprice} presents our tool-based solution, AutoPRICE. 
%We discuss the user study conducted on AutoPRICE in \Cref{US}, and 
We present ideas for future work in \Cref{discussion}, and discuss related work in \Cref{relatedwork}. 
We explain the limitations of our study in \Cref{limitations}, and conclude in \Cref{conclusion}.

%% file: Sections/Background.tex
\section{Background on Google's DSS}
\label{background}

%\mk{A very short section. Can we combine this with introduction?}

\begin{figure}[t]
\centering
\includegraphics[width=0.25\textwidth]{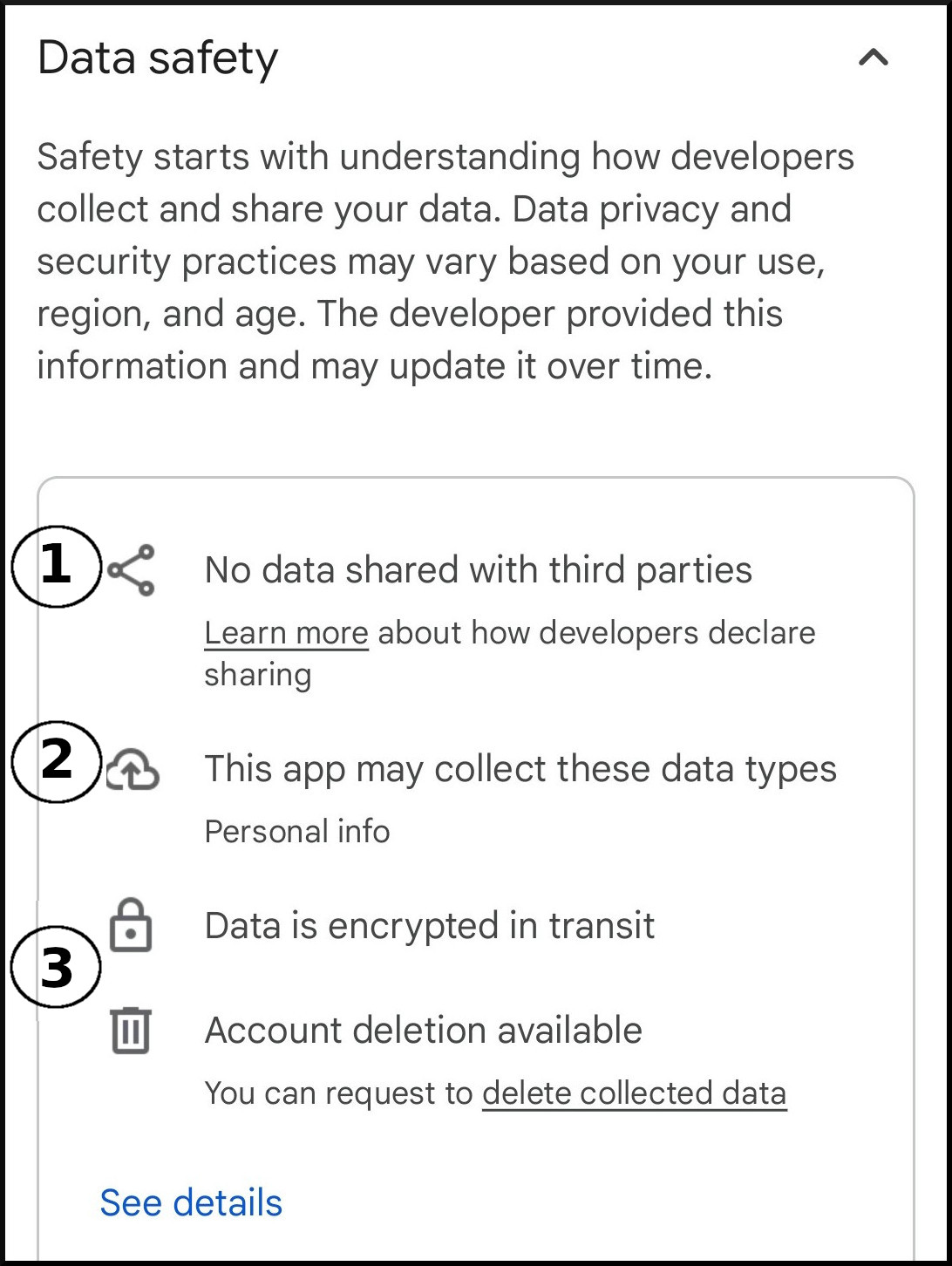}
\caption{Data Safety Section of Signal~\cite{signal}. \textcircled{1} Data sharing. \textcircled{2} Data collection. \textcircled{3} Security practices (encryption and data deletion).}
\label{fig:dss_1}
\end{figure}

\begin{figure}[t]
    \centering
         \includegraphics[width=0.25\textwidth]{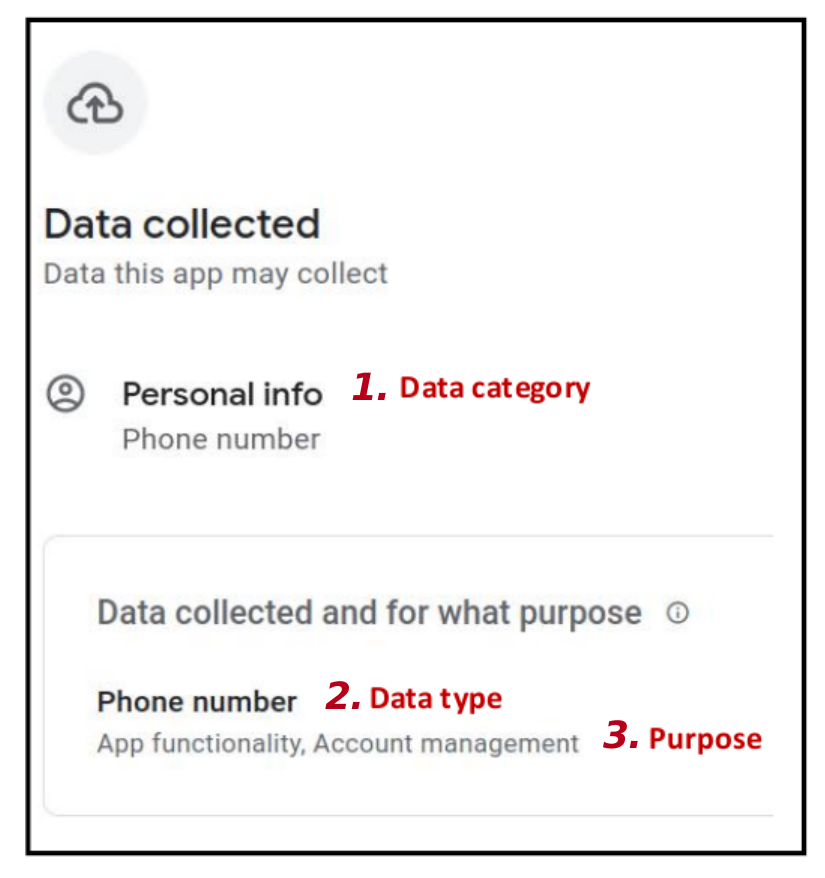}
         \caption{\textit{Data collected} part of the DSS of Signal~\cite{signal}. \stilltodo{\textit{1.}} Data category. \stilltodo{\textit{2.}} Data type. \stilltodo{\textit{3.}} Purpose.}
         \label{fig:dss_2}
\end{figure}

The DSS serves as a standardized disclosure mechanism, summarizing an app’s privacy practices in three main areas: \textit{data sharing}, \textit{data collection}, and \textit{security practices} (cf.~\Cref{fig:dss_1}). 
Within the DSS, data is classified into broad \textit{data categories} (e.g., location, personal information, financial information), which are further divided into specific \textit{data types} (cf.~\Cref{tab:DSScategories}). 
Developers must specify whether these data types are collected or shared, and for what \textit{purposes}, chosen from a predefined list. 
These disclosures are displayed on the app’s Play Store page, directly informing end-users about its privacy practices (cf.~\Cref{fig:dss_2}). 

\begin{table*}[t]
\centering
\small
\caption{Google Play's DSS data categories and data types~\cite{googledatatypes}.}
\label{tab:DSScategories}
\begin{tabular}{p{3.5cm} p{13cm}}
%\toprule
\rowcolor{gray!30} \textbf{Data Category} & \textbf{Data Types} \\
%\midrule
Location & Approximate location, Precise location \\
\rowcolor{gray!15} Personal Info & Name, Email address, User IDs, Address, Phone number, Race and ethnicity, Political or religious beliefs, Sexual orientation, Other info \\
Financial Info & Payment info, Purchase history, Credit score, Other financial info \\
\rowcolor{gray!15} Health \& Fitness & Health info, Fitness info \\
Messages & Emails, SMS/MMS, Other in-app or chat messages \\
\rowcolor{gray!15} Photos \& Videos & Photos, Videos \\
Audio Files & Voice or sound recordings, Music files, Other audio files \\
\rowcolor{gray!15} Files \& Docs & Files and documents\\
Calendar & Calendar events \\
\rowcolor{gray!15} Contacts & Contacts \\
App Activity & App interactions, In-app search history, Installed apps, Other user-generated content, Other actions \\
\rowcolor{gray!15} Web Browsing & Web Browsing history \\
App Info \& Performance & Crash logs, Diagnostics, Other performance data \\
\rowcolor{gray!15} Device or Other IDs & Device or other IDs \\
%\bottomrule
\end{tabular}
\end{table*}

To populate these disclosures, app developers are required to complete a DSS form in the Google Play Console~\cite{playconsole}. 
%An example of the DSS form is shown in \Cref{fig:dssform}. 
In this form, developers first select all relevant \textit{data types} their app collects or shares. 
For each selected type, they must provide detailed information, including:
\begin{itemize}
\item whether the data is collected, shared, or both;
\item whether it is processed ephemerally;
\item whether its collection is optional for users; and
\item the purposes for which the data is collected (selected from a fixed set).
\end{itemize}

Completing and submitting this form is mandatory for publishing any app on the Play Store, including apps that claim not to collect user data. 
Submitted forms are reviewed by Google as part of the app approval process to ensure compliance with the platform’s privacy reporting requirements.
\begin{comment}
\begin{figure}[t]
    \centering
         \includegraphics[width=0.45\textwidth]{Figures/DSSformexample.png}
         \caption{An example of the DSS form for the data type \emph{Email address}~\cite{dssformexample}.}
         \label{fig:dssform}
\end{figure}
\end{comment}

%% file: Sections/Methodology.tex
\section{Study Design}
\label{methodology_MMS}
 
We conducted an empirical study to achieve our goal discussed in \Cref{intro}. 
The study was designed to answer the following research questions:

\textit{\textbf{RQ1.} What methods and resources do developers use to categorize privacy-related data collected by their app into the DSS data categories?} 

This question focuses on understanding how developers currently approach data classification during the DSS completion process, and whether they use any tools or resources for this purpose. 
By examining these practices, we aim to identify areas where tooling or official documentation can be improved to better support developers in this task.

\textit{\textbf{RQ2.} How confident are developers in their ability to accurately and effectively complete the DSS forms?}

This question seeks to assess developers’ confidence in their ability to complete the DSS form accurately and in compliance with privacy requirements. 
Understanding developers’ confidence levels can reveal points of uncertainty or ambiguity in the reporting process, helping guide the design of more effective support tools and clearer guidance.

\textit{\textbf{RQ3.} What are the challenges developers face when completing the DSS forms?} 

This question focuses on the challenges developers face to complete the %\emph{data collected} part of the 
DSS forms. 
By identifying these challenges, we aim to uncover opportunities to improve developer support mechanisms and enhance the overall transparency and accuracy of privacy reporting practices. 

\begin{figure}[t]
\centering
\includegraphics[width=0.49\textwidth]{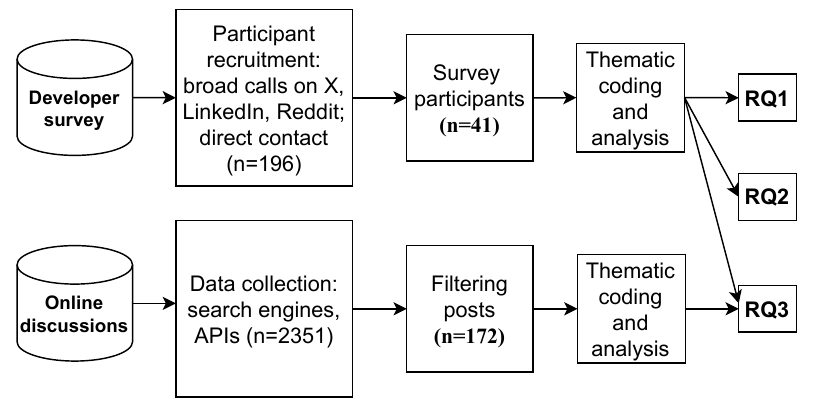}
\caption{Our study design.}
\label{fig:methodology}
\end{figure}

\Cref{fig:methodology} provides an overview of the study design. 
Our approach combines two complementary data sources: (1) a survey of Android developers capturing their self-reported practices and perceptions, and (2) a qualitative analysis of online developer discussions that reveal naturally occurring challenges in community settings. 
The survey directly addresses RQ1 and RQ2, while both data sources collectively answer RQ3 through thematic analysis.

%\eb{I find that the following is a bit low in detail.}

\subsection{Developer Survey} 
\label{survey}

To directly capture Android developers’ experiences with completing the DSS form, we conducted an online survey. 
Developer surveys are commonly used in software engineering research to generate discussions of subjects where individual experiences are critical~\cite{survey1, survey2, survey3, franke2024, survey4}.  
\begin{table}[]
\small
\begin{tabular}{clc}
\hline
\textbf{RQ}                                                                             & \multicolumn{1}{c}{\textbf{Question}}                                                                                                                                                          & \multicolumn{1}{l}{\textbf{\begin{tabular}[c]{@{}l@{}}Question \\ type\end{tabular}}}     \\ \hline
\multirow{2}{*}{\begin{tabular}[c]{@{}c@{}}RQ1: \\ methods \&\\ resources\end{tabular}} & \begin{tabular}[c]{@{}l@{}}How do you categorize the privacy-related \\ data collected by your app for the DSS?\end{tabular}                                                                    & \multirow{2}{*}{\begin{tabular}[c]{@{}c@{}}Multiple-\\ choice,\\ open-ended\end{tabular}}  \\ \cline{2-2}
                                                                                        & \begin{tabular}[c]{@{}l@{}}What resources do you use to help you \\ understand and complete the DSS?\end{tabular}                                                                               &                                                                                            \\ \hline
\multirow{5}{*}{\begin{tabular}[c]{@{}c@{}}RQ2: \\ confidence \\ levels\end{tabular}}   & \begin{tabular}[c]{@{}l@{}}How well do you understand the privacy-\\ related data your app collects \\ (e.g., location, personal, financial) \\ that must be disclosed in the DSS?\end{tabular} & \multirow{5}{*}{\begin{tabular}[c]{@{}c@{}}5-point \\ Likert \\ scale\end{tabular}}        \\ \cline{2-2}
                                                                                        & How familiar are you with GDPR?                                                                                                                                                                 &                                                                                            \\ \cline{2-2}
                                                                                        & \begin{tabular}[c]{@{}l@{}}How confident are you that your app’s \\ data collection in the DSS aligns \\ with GDPR requirements?\end{tabular}                                                   &                                                                                            \\ \cline{2-2}
                                                                                        & \begin{tabular}[c]{@{}l@{}}How well do you understand Google’s \\ definitions and categorizations of data \\ types within the DSS?\end{tabular}                                                 &                                                                                            \\ \cline{2-2}
                                                                                        & \begin{tabular}[c]{@{}l@{}}How challenging is it to fill out the Data \\ Collection part of the DSS?\end{tabular}                                                                               &                                                                                            \\ \hline
\multirow{6}{*}{\begin{tabular}[c]{@{}c@{}}\\RQ3: \\ challenges\end{tabular}}             & \begin{tabular}[c]{@{}l@{}}Which aspects of the DSS do you find \\ challenging?\end{tabular}                                                                                                    & \multirow{2}{*}{\begin{tabular}[c]{@{}c@{}}Multiple-\\ choice, \\ open-ended\end{tabular}} \\ \cline{2-2}
                                                                                        & \begin{tabular}[c]{@{}l@{}}What specific challenges have you faced \\ while filling out the DSS?\end{tabular}                                                                                   &                                                                                            \\ \cline{2-3} 
                                                                                        & \begin{tabular}[c]{@{}l@{}}Describe any technical or general issues \\ you have encountered during the \\ submission of the DSS.\end{tabular}                                                   & \multirow{2}{*}{Open-ended}                                                                \\ \cline{2-2}
                                                                                        & \begin{tabular}[c]{@{}l@{}}Have you ever had an app rejected due to \\ issues with the DSS? \\ If yes, please describe the situation.\end{tabular}                                              &                                                                                            \\ \hline
\end{tabular}
\caption{Mapping between research questions and survey questions.}% Full codebook available in \Cref{codebook}.}
\label{tab:rq-survey-codes}
\end{table}

\textbf{Survey Questionnaire.} We designed the questionnaire with a focus on the research questions. 
%\eb{you mean RQ1? Or also RQ2?}
%\mk{Both, clarified in the next sentence.} 
These items explored how developers identify privacy-related data collected by their apps (\textbf{RQ1}), how confident they are in completing the DSS form (\textbf{RQ2}), and the obstacles they face during this process (\textbf{RQ3}).  
The survey included 5-point Likert-type items, multiple-choice questions, and open-ended prompts. 
A compact version of the questionnaire is presented in \Cref{tab:rq-survey-codes}. 
Full questionnaire is available in the artifacts~\cite{artifacts}.
%The survey responses informed our analysis of the research questions.

\textbf{Participant Recruitment.} We distributed our survey in three rounds between August and December 2024. 
In the first round, we made broad calls for participation on X, LinkedIn, and Reddit. 
We received 19 responses, with only 4 respondents having apps published in the European Union. 
After a few months, we undertook the second round, during which the first author directly contacted over 150 randomly selected Android developers via LinkedIn. 
We received 15 responses, i.e., obtained a response rate of 10\% . 
In the third round, we contacted 46 additional app developers who were engaged in discussions about Android development, Google Play, and the DSS on GitHub and Reddit. 
We received 7 responses (15\% response rate). 
In total, we obtained data from 41 participants. %(19 from from initial broad outreach, 15 from direct LinkedIn contacts, and 7 from GitHub and Reddit). 

\textbf{Survey Participants.} %At the beginning of the survey, we asked questions regarding experience and demographics. 
Our participants have a median of 5 years of Android app development experience (average = 5.54). 
37 participants (90\%) confirmed that they have previously completed the DSS form, and 23 participants (56\%) have published app(s) in the EU. 
We share detailed demographics in \Cref{tab:surveyparticipants}. 
We refer to our survey participants as P01–P41 (cf.~\Cref{results_MMS}). %\stilltodo{TODO: remove table to save space.}

\begin{table}[t]
\centering
\small
\caption{Survey participants' demographics (based on optional responses).}
\label{tab:surveyparticipants}
%\begin{tabular}{l@{\hspace{5pt}}|l@{\hspace{5pt}}|l@{\hspace{5pt}}|l@{\hspace{5pt}}|l@{\hspace{5pt}}|l@{\hspace{5pt}}|l@{\hspace{5pt}}|l@{\hspace{5pt}}|l}
\begin{tabular}{lll}
\hline
%\textbf{Gender} & Male & 31 (71.6\%)\\
%& Female & 5 (12.19\%) \\
%\midrule
%\textbf{Age} & 25-34 & 30 (73.17\%)\\
%& 35-44 & 4 (9.75\%)\\
%& 45-54 & 2 (4.87\%) \\
%& 55-64 & 1 (2.43\%)\\
%\midrule
\textbf{Experience [years]} & No experience & 4 (9.75\%)\\
& Less than 1 year & 5 (12.19\%)\\
& 1-3 years & 8 (19.51\%)\\
& 4-6 years & 5 (12.19\%)\\
& 7-10 years & 10 (24.39\%)\\
& More than 10 years & 7 (17.07\%)\\
\hline
\textbf{Published app in EU ($\ast$)} & Yes & 23 (56\%)\\
& No & 18 (43.9\%)\\
\hline
\end{tabular}
\end{table}

\textbf{Thematic Coding and Analysis.} We used an iterative open coding approach~\cite{opencoding} to perform thematic analysis~\cite{thematic} of open-ended survey responses. 
The first author carried out inductive coding to derive emergent themes, which were then compiled into a codebook. 
The second author independently reviewed the annotations, and disagreements were discussed and resolved through refinement of the final coding scheme. 
During this discussion, we refined 5 subthemes, resulting in re-alignment of 1 code (out of 24 codes).  
The final codebook is available in \Cref{codebook}. 
%A single researcher (the first author) conducted all of the coding using an inductive process where major themes emerge through interpretation of the data. 
%This resulted in a codebook, which was double-checked by the second author. 
%Both researchers reached partial agreement, after which the codebook was modified.

\subsection{Analysis of Developer Discussions} 
\label{discussions}

\begin{table}[!htb]
\centering
\small
\begin{tabular}{l l r r r}
\hline
\textbf{\begin{tabular}[c]{@{}l@{}}Online \\Platform\end{tabular}} & \textbf{\begin{tabular}[c]{@{}l@{}}Extraction\\ Source(s)\end{tabular}} & \textbf{\begin{tabular}[c]{@{}l@{}}Raw\\posts\end{tabular}} & \textbf{\begin{tabular}[c]{@{}l@{}}Final\\posts\end{tabular}} & \textbf{\begin{tabular}[c]{@{}l@{}}Unique \\Developers\\(Final)\end{tabular}} \\ \hline
Stack Overflow & API + Google & 1505 & 56 & 268 \\ \hline
Reddit & Google & 235 & 65 & 283 \\ \hline
Discord & Local search & 29 & 28 & 31 \\ \hline
GitHub & GitHub + Google & 572 & 22 & 48 \\ \hline
Hacker News & Google & 10 & 1 & 12 \\ \hline
\textbf{Total} & --- & \textbf{2351} & \textbf{172} & \textbf{642} \\ \hline
\end{tabular}
\caption{Summary of developer discussions collected across platforms. “Raw posts” refers to all extracted discussions (may include duplicates and irrelevant posts), and “Final posts” refers to those retained after filtering.}
\label{tab:posts-devs}
\end{table}

Our analysis of the survey data revealed that many participants rely on online forums and developer communities as key resources when completing the DSS form (see \Cref{rq1}). 
To capture a broader range of perspectives and enable data triangulation~\cite{triangulation}, we therefore complemented the survey with a qualitative analysis of online developer discussions.

\textbf{Data Collection.} We used literature~\cite{developercommunities,reddit1,reddit2, SOreddit, SOreddit1, survey3} to systematically identify relevant public forums and developer communities where Android privacy and data reporting topics are actively discussed. 
We also conducted pilot searches to refine our choice of sources and search terms. 
Our search focused on popular communities such as r/android (3.1M members), r/androidapps (436K), and r/androiddev (261K) on Reddit, as well as r/AndroidDev (34.5K) on Discord.

To capture posts relevant to the DSS form, we experimented with several keyword combinations and finalized the following query:
%After a pilot search with several term combinations, we finalized the following query: 
(\textit{DSS OR ``data safety section'' OR ``data safety form''}).  
Using this query, we collected posts from multiple developer platforms (cf.~\Cref{tab:posts-devs}). 
We leveraged both Google and local search engines, and used the Stack Exchange API~\cite{stackapi} to automatically extract posts tagged with relevant keywords. 
Data collection took place between January and September 2025, resulting in 2,351 discussions in total. 
These included potential duplicates and irrelevant discussions. 
Further details on data collection including search queries and Python scripts using developer community APIs are provided in the accompanying artifacts~\cite{artifacts}. 

\textbf{Filtering Posts.} We applied a two-stage filtering process to ensure that only discussions authored by app developers and those directly related to the DSS were retained. 
Posts were included if they explicitly discussed experiences with DSS completion, privacy reporting, or app rejections related to the DSS or privacy policies. 
Posts were excluded if they were unrelated to the DSS reporting process (e.g., news articles or general privacy discussions) or authored by non-developers (e.g., lawyers or users commenting on app privacy). 
For Stack Overflow, we used the Stack Exchange API to further filter out posts discussing unrelated platforms (e.g., iOS or Apple labels). 
After filtering, our final dataset comprised 56 Stack Overflow posts, 65 Reddit threads, 28 Discord discussions, 22 GitHub issues and discussions, 1 post from Hacker News~\cite{hackernews}, all posted after the DSS was introduced in 2022. 

\textbf{Measuring Engagement.} To measure user engagement, we used Stack Exchange API~\cite{stackapi}, Reddit API~\cite{redditapi}, and GitHub API~\cite{githubapi} to automatically count unique developers interacting with the posts. 
For Discord, participant counts were obtained manually, while for Hacker News, we used a custom Python script to identify unique commenters. 
Altogether, our dataset comprised 172 posts, involving 642 distinct developers (cf.~\Cref{tab:posts-devs}). 

\textbf{Thematic Coding and Analysis.} We followed the same process of qualitative analysis as the developer survey. 

%% file: Sections/Results.tex
\section{Study Results}
\label{results_MMS}

\begin{figure*}[t]
\begin{subfigure}[b]{0.5\textwidth}
\includegraphics[width=\textwidth]{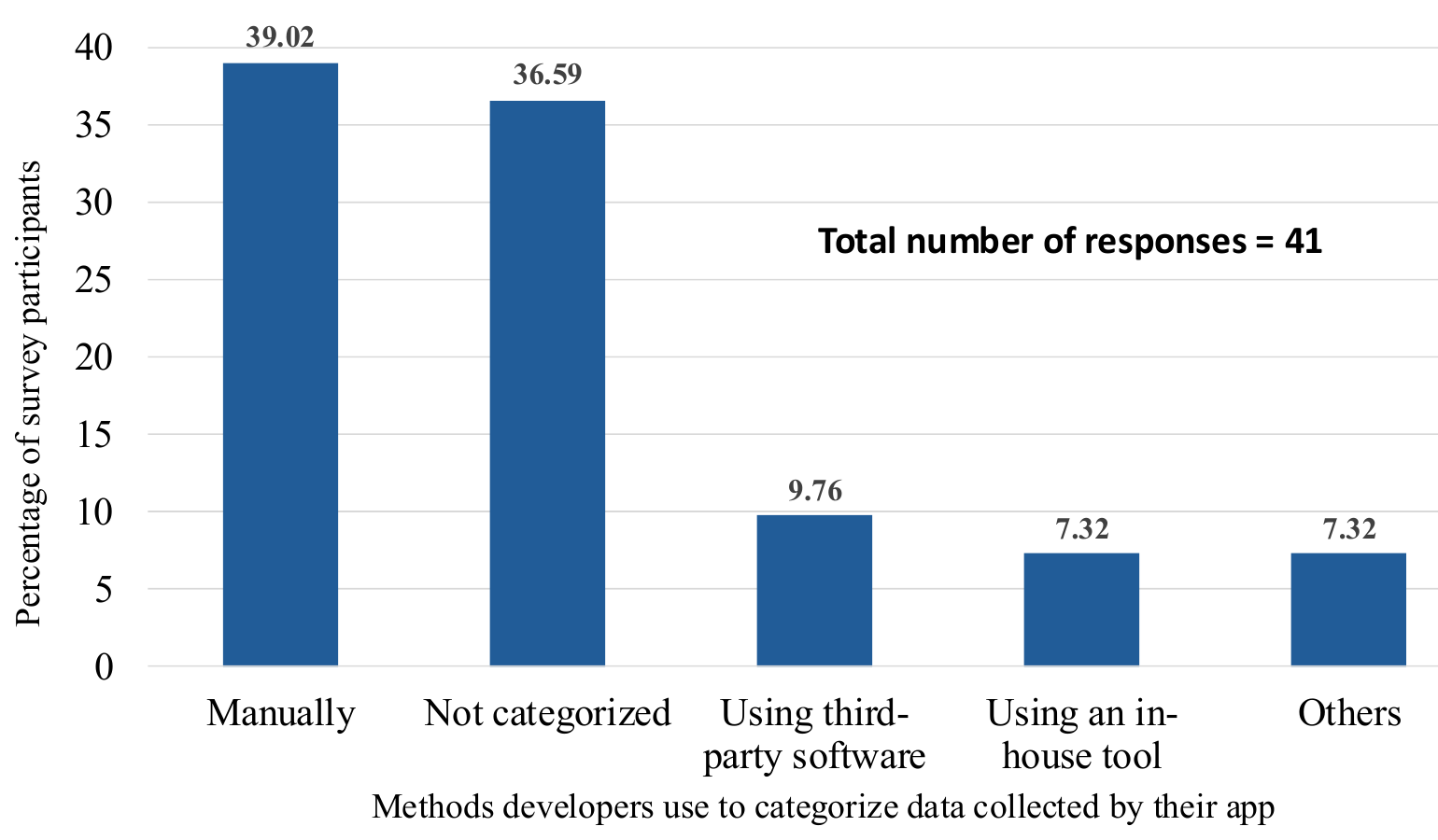}
\caption{Bar chart illustrating what methods survey participants use categorize data collected by their apps.}
\label{fig:howtocategorize}
\end{subfigure}
\hspace{0.5cm}
\begin{subfigure}[b]{0.45\textwidth}
\includegraphics[width=\textwidth]{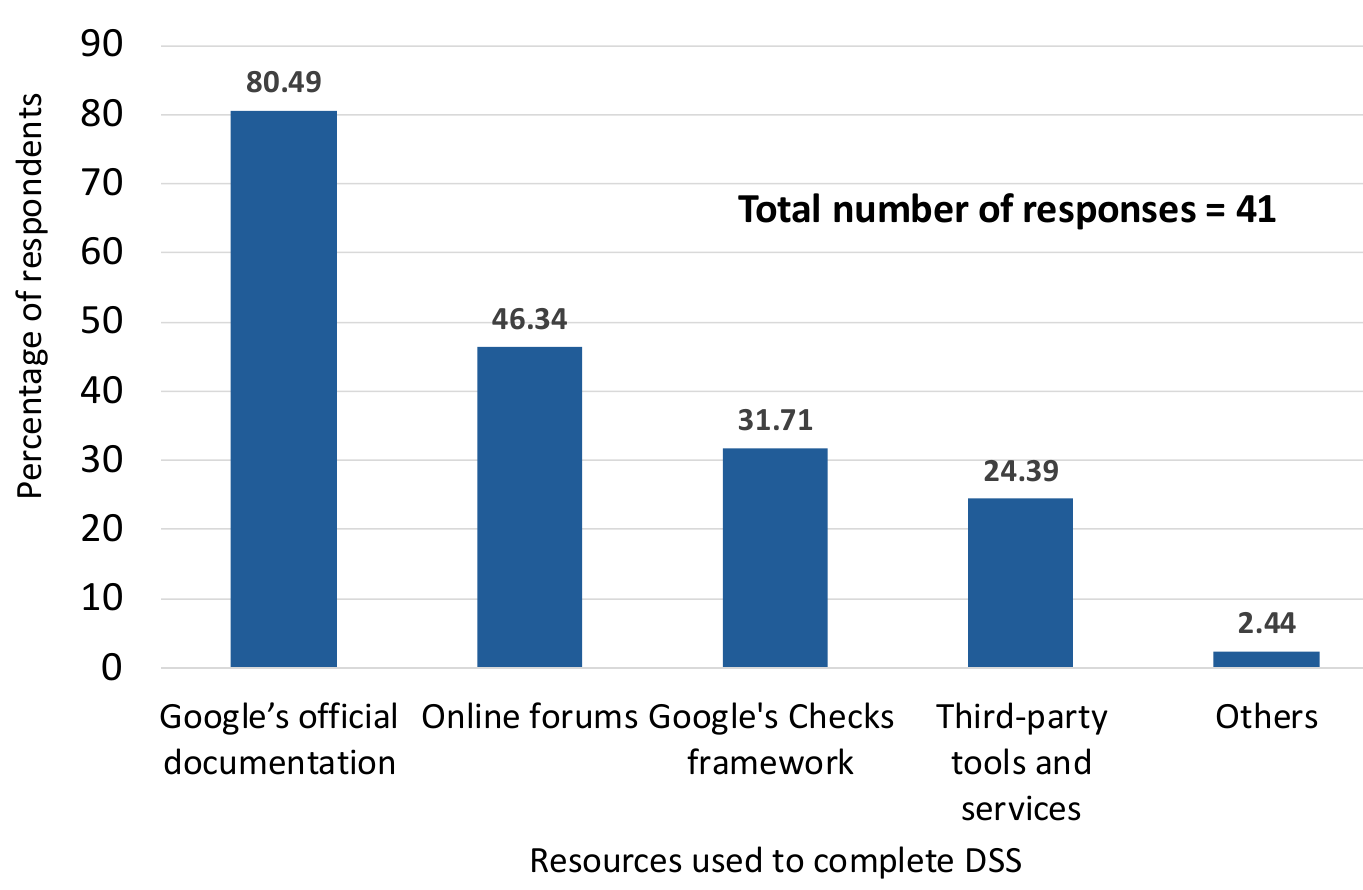}
\caption{Bar chart showing the percentage of participants who used resources to complete the DSS form. Participants could choose multiple options for this question.}
\label{fig:resources}
\end{subfigure}
\centering
\caption{Methods and resources participants use to categorize privacy-related data collected by their app into the DSS data categories (RQ1).}
\end{figure*}

We discuss the results of the developer survey in the context of all three research questions presented in \Cref{methodology_MMS}, and analysis of developer discussions in the context of the third research question. 

\subsection{\textbf{RQ1: Methods and Resources}}
\label{rq1}

The DSS form requires developers to disclose collected information under broad categories (e.g., location, personal information, financial information) that are further divided into specific data types (cf.~\Cref{tab:DSScategories}). 
We asked our survey participants what methods they use to classify the privacy-related data collected by their apps in accordance with these categories. 
The results are shown in \Cref{fig:howtocategorize}. 
\textbf{39.02\%} participants reported classifying data manually. 
P03 explained, \directquote{The data is categorized manually to ensure accuracy and compliance with Google Play’s guidelines, allowing for thorough review and customization based on the app’s specific data collection practices.} 

However, more than a third of participants (\textbf{36.59\%}) indicated that they do not perform any data categorization. 
In practice, this could mean that they select all data categories as collected, select none, or just guess which categories apply for their app. 
Each of these behaviors could result in inaccurate or incomplete disclosure, giving end users a false sense of privacy about how their data is handled. 
Several participants attributed this to the perception that their apps did not collect personal data. 
For instance, P22 said, \directquote{I did not need to store any customer related data.} 
P24 said, \directquote{(I am) not collecting any data.} 
P20 added a pragmatic perspective: \directquote{I try to avoid doing that (completing the DSS form) as much as possible. Usually just by not collecting any data. It makes the user experience worse but affordable.} 
Others, such as P16, emphasized the simplicity of their applications: \directquote{My app is simple, so no need to care about this (DSS form).} 

Only \textbf{9.76\%} reported relying on third-party solutions for categorization, highlighting the limited availability of automated, context-aware approaches. 
P41 noted, \directquote{Many developers lack tools to automatically track and audit data usage, making it challenging to identify all collected or shared data.} 
A smaller group (\textbf{7.32\%}) reported using in-house tools and other approaches, such as permissions, to support categorization. 
P20 explained, \directquote{App permissions (runtime or manifest-based) provide the most accurate info.} 

We next inquired about all the resources participants rely on to complete the DSS form (cf.~\Cref{fig:resources}). 
Most participants rely on Google’s official documentation (\textbf{80.49\%}), followed by online forums and communities (\textbf{46.34\%}). 
Other reported resources include Google’s Checks framework (\textbf{31.71\%}) and third-party services (\textbf{24.39\%}). 
P39 used \textit{Privado}~\cite{privadoai} to complete the DSS form. \\

\begin{mdframed}[backgroundcolor=black!10,roundcorner=8pt]
\textbf{\textbf{Finding 1.}} Developers use existing online resources to complete the DSS form. Most either manually categorize the collected data or omit categorization altogether.
\end{mdframed}

\subsection{\textbf{RQ2: Confidence Levels}}
\label{rq2}

\begin{figure*}[!t]
%\begin{center}
\includegraphics[width=0.7\textwidth]{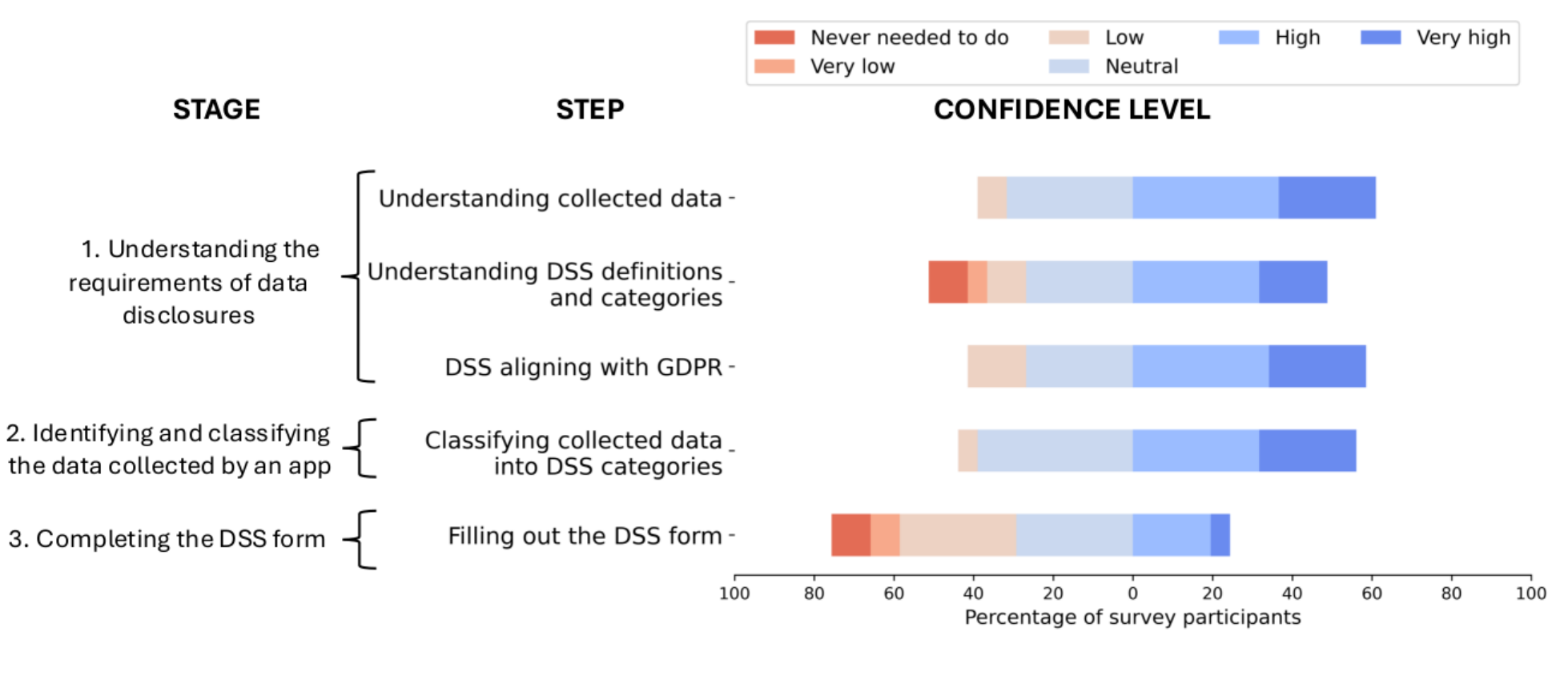}
\caption{Survey participants' responses to the questions about confidence levels for different processes (RQ2).}
\label{fig:confidencelevels}
%\end{center}
\end{figure*}

To investigate this question, we analyzed the survey responses. % and online developer discussions. 

We sought to identify \textit{where} in the process of privacy-aware reporting developers encounter the most difficulty. 
We conceptualize this process as consisting of three stages: 
\begin{enumerate}
    \item[1.] understanding the requirements of data disclosures,
    \item[2.] identifying and classifying the data collected by an app, and 
    \item[3.] completing the DSS form.
\end{enumerate}

We asked survey participants to rate their confidence in these stages. 
The results are shown in \Cref{fig:confidencelevels}.  
While many developers reported confidence in understanding and classifying privacy-related data (stages 1 and 2), their confidence dropped noticeably when it came to accurately completing the DSS form (stage 3.). 

For instance, P03 noted confidence in identifying and categorizing data but highlighted the difficulties of nuanced distinctions: \directquote{I have a good understanding of Google's definitions and categorizations of data types within the DSS. However, some nuances, such as differentiating between data types that seem similar or ensuring compliance with the latest updates to the guidelines, can occasionally be confusing and require further clarification or consultation with relevant documentation.} 
They further explained: \directquote{Challenges may arise when distinguishing between data categories like precise versus approximate location, especially when integrating third-party services or handling edge cases. Regular reviews and consultations with documentation help mitigate these challenges.} 

Other participants expressed more uncertainty. 
P27, for example, admitted: \directquote{When I use third-party libraries, I don't have full control over what's collected.} 
Similarly, P09 observed broader inconsistencies across contexts: \directquote{Country-wise app data collection rules are different. And there are tons of apps which don't comply with these rules and still available on Play Store.} 

Overall, these perspectives reveal a clear pattern: developers are generally confident in recognizing the data their apps collect, yet they face challenges in translating this knowledge into DSS-compliant disclosures. 
Even with familiarity with Google’s data type definitions~\cite{googledatatypes}, uncertainty arises when handling nuanced categories, third-party services, or varying regulatory expectations. \\

\begin{mdframed}[backgroundcolor=black!10,roundcorner=8pt]
\textbf{\textbf{Finding 2.}} Developers are generally confident in identifying the data their apps collect, yet their confidence drops significantly when translating this knowledge into DSS-compliant disclosures. 
\end{mdframed} 

\subsection{\textbf{RQ3: Challenges}}
\label{rq3}

\begin{figure}[t]
\includegraphics[width=0.5\textwidth]{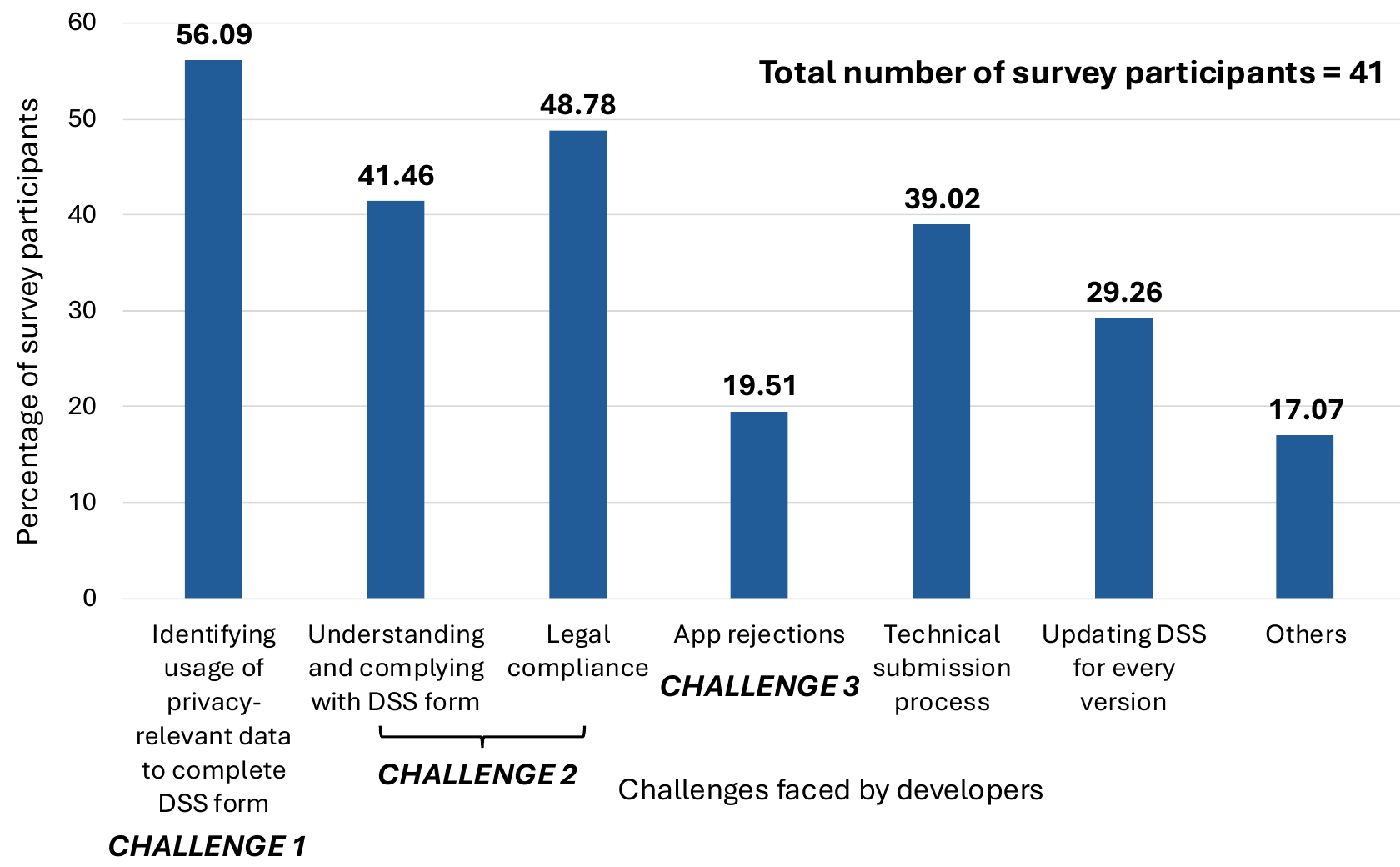}
\caption{Bar chart showing the distribution of challenges across survey participants (RQ3). Participants could choose multiple challenges through various questions.}
\label{fig:challenges}
\end{figure}

To investigate the \emph{specific} challenges developers face when completing the DSS form, we drew upon two complementary sources: survey answers and online developer discussions.  
For the survey, we analyzed responses to all questions related to RQ3 (cf.~\Cref{tab:rq-survey-codes}), which provided qualitative insights into developers’ experiences. 
These included questions where participants ranked predefined challenges identified from existing literature~\cite{datalabels,khedkar2024androidappdevelopersaccurately}, and optionally added their own challenges under an ``Others'' field or in other open-ended questions. 
We coded the survey responses to determine how many developers encountered each challenge. 
The resulting categorization is summarized in \Cref{fig:challenges}. 

We further examined online discussions from Stack Overflow, Reddit, Discord, GitHub, and Hacker News (posted after the DSS was introduced in 2022). 
The overall distribution of challenges identified across these online discussions is shown in \Cref{tab:posts-devs-detail}. 
We categorized these discussions at two levels of granularity. 
At the first level, we grouped these discussions  according to the high-level challenge they addressed (e.g., \emph{understanding and complying with DSS form}). 
At the second level, we further subdivided each challenge into specific subtopics that captured recurring themes within the discussions (e.g., within \emph{understanding and complying with DSS form}, discussions were split into \emph{confusing terminologies}, \emph{data deletion}, \emph{encryption declaration}, and \emph{legal compliance}). 

While we applied a similar categorization process to the survey data, the online discussions provided a more comprehensive taxonomy of challenges due to their greater volume, diversity of perspectives, and depth of qualitative detail.  
This hierarchical analysis enabled us to identify not only the major categories of developer difficulties but also nuanced variations within each challenge.  

\begin{table*}[ht]
\small
\centering
\rowcolors{2}{gray!10}{white}
\setlength{\tabcolsep}{3pt} % tighter columns
\begin{tabular}{lccccccccc}
\rowcolor{gray!30}
\textbf{Data Sources} &
\multicolumn{3}{c}{\textbf{\begin{tabular}[c]{@{}c@{}}Challenge 1: Identifying usage of\\
privacy-relevant data to complete \\DSS form\end{tabular}}} &
\multicolumn{4}{c}{\textbf{\begin{tabular}[c]{@{}c@{}}Challenge 2: Understanding and complying \\with DSS form\end{tabular}}} &
\textbf{\begin{tabular}[c]{@{}c@{}}Challenge 3: \\Understanding\\ app \\rejections\end{tabular}} &
\textbf{\begin{tabular}[c]{@{}c@{}}Other\\challenges\end{tabular}} \\

\rowcolor{gray!30}
& \textbf{a) TPL data} & \textbf{b) Data} & \textbf{c) Data} & 
\textbf{a) Confusing} & \textbf{b) Data} & \textbf{c) Encryption} & \textbf{d) Legal} &
 & \\
\rowcolor{gray!30}
& \textbf{collection} & \textbf{collection} & \textbf{sharing} &
\textbf{terminologies} & \textbf{deletion} & \textbf{declaration} & \textbf{compliance} &
 & \\
Stack Overflow & 15 & 21 & 0 & 1 & 0 & 4 & 0 & 14 & 1 \\
Reddit & 28 & 11 & 3 & 10 & 9 & 1 & 0 & 3 & 0 \\
Discord & 11 & 8 & 0 & 4 & 1 & 0 & 0 & 4 & 0 \\
GitHub & 11 & 1 & 3 & 2 & 1 & 0 & 0 & 2 & 2 \\
Hacker News & 0 & 0 & 0 & 0 & 0 & 0 & 0 & 1 & 0 \\  \hline
\begin{tabular}[c]{@{}l@{}}Total across \\platforms (n=172)\end{tabular} & \textbf{65} & \textbf{41} & \textbf{6} & \textbf{17} & \textbf{11} & \textbf{5} & \textbf{0} & \textbf{24} & \textbf{3} \\ 
\textbf{\begin{tabular}[c]{@{}l@{}}Percentage \\(out of 172)\end{tabular}} & \textbf{37.79\%} & \textbf{23.83\%} & \textbf{3.48\%} & \textbf{9.88\%} & \textbf{6.39\%} & \textbf{2.9\%} & \textbf{0\%} & \textbf{13.95\%} & \textbf{1.74\%} \\ \hline

%\begin{tabular}[c]{@{}l@{}}Open-ended survey \\answers (n=44)\end{tabular} & 12 & 9 & 0 & 13 & 0 & 0 & 6 & 8 & 3 \\ \hline
%\textbf{\begin{tabular}[c]{@{}l@{}}Percentage \\(out of 44)\end{tabular}} & \textbf{27.3\%} & \textbf{20.5\%} & \textbf{0\%} & \textbf{29.5\%} & \textbf{0\%} & \textbf{0\%} & \textbf{13.63\%} & \textbf{18.2\%} & \textbf{6.81\%} \\
%\hline   
\end{tabular}
\caption{Distribution of challenges across online developer discussions (RQ3). At the first level, discussions are grouped according to the high-level challenge they addressed. Each challenge is further subdivided into specific subtopics that captured recurring themes. Each online discussion is mapped to a single subtopic. Abbreviation: TPL = third-party library.}
\label{tab:posts-devs-detail}
\end{table*}
The following provides a detailed description of the recurring challenges observed: 
\subsubsection{Identifying usage of privacy-relevant data to complete DSS form.}

Developers are required to populate the \emph{data collected} or \emph{data shared} sections of the DSS form (\textcircled{1} and \textcircled{2} in \Cref{fig:dss_1}). 
Many reported uncertainty regarding which types of data in their apps should be considered privacy-relevant and disclosed in these sections. 
Specifically, \textbf{56.09\%} of survey respondents (cf.~\Cref{fig:challenges}) and \textbf{61.62\%} of online discussions (106 out of 172) focused on this issue. 

\textbf{a) Third-Party Library (TPL) data collection.} One of the most frequently reported issues was the lack of clarity and control over data collected by third-party libraries. 
This concern appeared in 65 online discussions (37.79\%) and was mentioned by 7 survey participants (17.07\%). 
Developers frequently sought community guidance on how to report data collected by SDKs integrated into their apps, such as analytics, login, or advertising libraries\footnote{An example (TPL data collection): \url{https://www.reddit.com/r/androiddev/comments/s7mk24/safety_data_admob/}}. 
The most frequently cited third-party libraries were Admob~\cite{admob} (21 mentions), Firebase~\cite{firebase} (19), Unity~\cite{unity} (6), Google services (Maps, Billing, Play Services; 3 each), and AppLovin~\cite{applovin} (3). 

Survey participants emphasized the difficulty of identifying data collected by third-party code and distinguishing it from data collected directly by the app. 
As P39 explained, \directquote{When third-party libraries are involved, we would need to declare data collected by those libraries too. 
In the current situation, it appears as if the app itself collects that information. 
The reality is a bit different. 
It might be better if there is some way to clearly communicate this distinction.} 
P33 and P40 highlighted the challenges of determining what data is collected through external SDKs and the lack of accurate third-party documentation. 
Some developers, such as P35, recommended using only SDKs whose behavior could be fully verified, ideally by reviewing source code. 
Such issues also resulted in app rejections. 
For instance, P27 recalled, \directquote{I developed an app that uses Google Maps. When I uploaded a release, it was rejected for not declaring that my app uses a library (Google Maps) that collects user location.}

%P33 highlighted the difficulty in determining what data is collected when \textit{it is code that we have little control over.} 
%P40 said, \directquote{Accurately identifying and categorizing all types of user data collected by the app (e.g., location, personal information, device data) can be challenging, especially in apps with third-party SDKs or APIs.} 
%They further criticized the \textit{inaccuracy of third-party documentation}.  
%To address these challenges, P35 recommended only using third-party SDKs when their behavior is fully understood—ideally by reviewing their source code. 

\textbf{b) Data collection.} Challenges were not limited to data collected by third-party libraries. 
41 online discussions (23.83\%) and 7 survey participants (17.07\%) were unsure if a certain data item needs to be disclosed in the data collected part of the DSS form. 

13 Stack Overflow posts\footnote{An example (data collection error): \url{https://stackoverflow.com/questions/71199143/}} discussed the following Play Store error messages: \emph{we detected user data transmitted off device that you have not disclosed in your app's Data safety form as user data collected.} 
Other discussions revealed confusion over what qualifies as data collection—such as metadata embedded in images (e.g., location), crash reports, Application Not Responding (ANR) reports, SDK activity, or login forms—and whether it needs to be disclosed. 
Some others struggled to identify the collected data items requiring mandatory disclosure via the DSS form. 
%Some others struggled with edge cases such as whether login email addresses must be declared as personal identifiers.

Among the survey responses, participants reported difficulties when apps acted primarily as frontends for remote services. 
As P39 explained, \directquote{The manual work required is a bit challenging. I'm always a bit skeptical as to whether I've accurately captured the data safety details in Play store. One specific challenge that comes to my mind is when a remote server is involved in the mix. Our app primarily acts a frontend to a remote service. It is unclear whether we need to point out the data collected by the service in Play Store or not. This is not clear as the user has to sign up separately for the service and would've already agreed to their terms and policies. So, it is not clear whether we need to reiterate this.}

\textbf{c) Data sharing.} 6 online discussions (3.48\%) focused on understanding if certain collected data needs to be disclosed as shared. 
These discussions highlighted confusion about the data sharing practices of Google SDKs, Firebase, AdMob, and notifee~\cite{notifee}. 
No survey respondents mentioned this issue. 

\subsubsection{Understanding and complying with DSS form.} 

Developers expressed frustration in understanding the DSS form itself. 
\textbf{58.53\%} of survey participants (cf.~\Cref{fig:challenges}) and \textbf{19.18\%} of discussions marked understanding and complying with DSS as a major challenge. 

\textbf{a) Confusing terminologies.} 16 online discussions (9.88\%) and 8 survey participants (19.51\%) highlighted confusion in understanding the DSS terminologies and the structure of the form. 

A user from an online discussion said, \directquote{There are so many questions, and I have no idea what to answer.} 
%Community advice varied widely. 
%%One user recommended, \directquote{Just say yes everywhere in the form,} while another suggested, \directquote{Minimize the number of fields—most apps do that. 
%Trying to be too thorough may lead to discrepancies between your Play Store entry and your privacy policy.} 
%Others dismissed the form: \directquote{It’s just a formality, like cookie disclaimers—just check what’s mentioned in the complaint.}  
Some developers struggled to interpret terminologies such as ephemeral data processing, and understand the differences between the definitions of data collection and sharing. 
Some others expressed confusion on what is considered a required data type\footnote{An example (confusing terminologies): \url{https://www.reddit.com/r/androiddev/comments/rf5b8u/data_safety_required_and_optional_data_collection/}}, and if accounts created by third-party SDK also need to be disclosed under data collection and sharing. 

Similar sentiments appeared in the survey answers. 
P20 admitted, \directquote{To be honest, I have no idea about any of the DSS data types, this is a mess. (I don't know) how to write it and not spend years on that, especially, to meet some uncertain requirements from Google (aka gray zone rules).}  
P34 noted that the wording of terms was unclear, and P27 emphasized the burden of navigating numerous categories and subcategories to avoid violations. 
P12 found the DSS form lengthy. 
P37 complained about excessive buttons and questions, explaining, \directquote{I had to create a bot to automate the process for my apps.} 
P31 described the effort of carefully parsing each question as a challenge in itself. 

\textbf{b) Data deletion.} 11 discussions (6.39\%) addressed challenges related to Google Play’s data deletion requirements (\textcircled{3} in \Cref{fig:dss_1}). 
Developers described confusion around implementing deletion mechanisms in apps that did not collect personal data or maintain user accounts. 
Others reported difficulties generating a data deletion URL when data was only collected through third-party SDKs. No survey respondents mentioned this issue. 

\textbf{c) Encryption declaration.} Only five online discussions (2.9\%) focused on issues related to declaring data encryption in the DSS form (\textcircled{3} in \Cref{fig:dss_1}). 
These developers reported receiving error messages from Google such as: \emph{You have declared that user data is encrypted in transit in your app’s Data Safety form and we’ve detected unencrypted network traffic that may carry user data off device.} 
In most cases, developers attributed these errors to third-party library dependencies but were unable to identify which specific dependency triggered the violation. 
No such challenges were reported in the survey responses. 

\textbf{d) Legal compliance.} 20 survey participants (48.78\%) found it challenging to comply with legal regulations and Google's policies. 
P13 observed \emph{compliance with multiple regulations} to be a major challenge. 
P34 noted, \directquote{There are several relevant Google Play policies, and these policies can change, difficult to check all the policy details}. 
No online discussions addressed this issue, which was unsurprising given that such discussions typically focus on seeking developer feedback.  

\subsubsection{Understanding app rejections.} Several challenges identified in our analysis were directly linked to app rejections from the Google Play Store. 
While rejections related to specific issues—such as identifying data usage and complying with the DSS—are discussed in their respective sections, here we focus on other app rejection experiences reported by 24 online discussions (13.95\%) and 8 survey participants (19.51\%).

Across online discussions, the most frequently cited reason for app rejections was a mismatch between the information provided in the DSS and the app’s privacy policy\footnote{An example (app rejection): \url{https://stackoverflow.com/questions/64929059/}}. 
Even developers who used automated tools such as Privado~\cite{privadoai} reported repeated rejections due to difficulties in accurately reporting data practices. 

Among the survey responses, P03 said, \directquote{My app was rejected due to issues with the DSS because the submitted information was either incomplete or inaccurately described. 
This included miscategorizing data types, not fully aligning with Google Play's privacy guidelines, or discrepancies between the disclosed data practices and the actual app functionality. 
Additionally, technical errors or non-compliance with regulations like GDPR might have contributed to the rejection.} 
P24 recalled, \directquote{(I was) getting threatened with having my app removed for not providing a data removal procedure, even though there is no data to be removed.} 
P34 added, \directquote{A feature that we added a long time ago to our app (in 2016) and that we forgot to mention in the DSS (caused the rejection).}

\subsubsection{Other challenges.}  
Beyond these challenges, 16 survey participants (39.02\%) identified issues with technical submission processes to be a major challenge, while 12 participants (29.26\%) emphasized difficulties in updating the DSS for each new app version. 
In addition, 1.74\% of online discussions and 17.07\% of survey participants discussed other challenges such as lack of tool support and missing ``security review'' within the DSS for their apps. \\

\begin{mdframed}[backgroundcolor=black!10,roundcorner=8pt]
\textbf{\textbf{Finding 3.}} Developers face many challenges when completing the DSS forms, including issues in identifying privacy-relevant data to complete the form, limited understanding of the form, and concerns about app rejection due to discrepancies with Google’s privacy requirements. 
\end{mdframed}

%% file: Sections/Discussion.tex
\section{Discussion}
\label{discussion}

In this section, we discuss our findings across four themes: (1) current practices of completing the DSS form, (2) challenges in reporting data collection, (3) the need to clarify the DSS form’s requirements, and (4) the role and limitations of existing tools. 

\textbf{Current Practices.} Developers mostly rely on online guidance to complete the DSS form, either manually classifying data or omitting classification altogether (\textbf{Finding 1}). 
In practice, this often means selecting all data categories as collected, none at all, or guessing which ones apply. 
This risks misreporting, leading to inaccurate disclosures and giving users a false sense of privacy. 

Several developers reported \emph{avoiding data collection altogether} to bypass the DSS process. 
As P20 explained, \directquote{I try to avoid doing that [completing the DSS form] as much as possible. Usually just by not collecting any data. It makes the user experience worse but affordable.} 
This reflects a pragmatic yet concerning trade-off, where functionality is reduced to escape compliance overhead. 
%Such practices underscore the perception that DSS compliance is burdensome and disconnected from practical development workflows.

While developers generally feel confident in identifying the data their apps collect (\textbf{Finding 2}), their confidence drops when translating this knowledge into DSS-compliant disclosures. 
This gap highlights that the challenge lies not in data awareness, but in interpreting Google’s disclosure requirements.

%Moreover, the DSS form is often completed by compliance teams rather than developers themselves. 
In some cases, it remains unclear whether app developers are even the ones filling out the form.
As P29 noted, \directquote{The release onus lies on the compliance team,} while another developer declined survey participation, noting, \directquote{Our compliance team does all law-paper stuff manually for all cases and platforms.} 
Delegation to compliance teams indicates a disconnect between the DSS’s intent—holding developers accountable—and its real-world execution.

To mitigate this gap, we recommend joint review processes involving both developers and compliance teams to align technical understanding with formal DSS reporting.

\textbf{Reporting Data Collection.} A key challenge highlighted in our study was reporting \emph{collected data} through the DSS form. 
Although the developer survey included some questions specifically about data collection (\textbf{RQ1}), our analysis of online discussions that examined developer challenges more broadly (\textbf{RQ3}) revealed that data collection remained the \emph{most prominent issue}: difficulties in reporting it, and the resulting errors and app rejections, were discussed in 106 out of 172 cases (61.62\%). 

A key factor behind this challenge is developers’ limited insight into the data gathered by \emph{third-party SDKs}. 
Although some SDK providers offer guidance pages instructing app developers what data types their SDKs use and thus must be declared to the DSS, a recent study~\cite{guidancepages} has detected inconsistencies between the guidance pages and the actual data collection of SDKs. 
Similarly, P33 noticed, \directquote{Data collected by third-party SDKs is not listed or explained clearly in their documentation.} 
This lack of reliable, actionable guidance leaves developers without adequate support. 

We recommend tools that detect SDK data collection and provide actionable disclosure suggestions, supported by standardized, machine-readable metadata and clearer Google DSS guidelines.

%we advocate for the development of integrated tools that can automatically detect data collected by SDKs and provide developers with actionable reporting suggestions. 
%Such tools should be complemented by standardized disclosure guidelines maintained by Google to ensure consistency across SDKs. 
%Additionally, requiring SDK providers to supply verified, machine-readable metadata about their data collection practices could give developers greater transparency and reduce reliance on unclear documentation. 
%Together, these measures would provide developers with both visibility and support, ultimately leading to more accurate DSS reporting and fewer app rejections. 
%There is a pressing need for focused tooling in this area. 
%While AutoPRICE takes a meaningful first step by detecting data collection by third-party libraries

\textbf{Clarifying the DSS Form.} A major challenge lies in the ambiguity of DSS terminology and reporting requirements. 
Many developers expressed uncertainty about \emph{what} information to disclose and \emph{why}. 
Our study identifies several areas needing clearer guidance: (1) vague definitions of ephemeral data processing (cf.~\Cref{background}), (2) ambiguity in data type categories (cf.~\Cref{tab:DSScategories}) (e.g., classifying video recordings without gallery access), (3) unclear account creation criteria involving third-party SDKs, (4) confusion over exemptions and required data types, and (5) difficulty distinguishing data collection from data sharing. 
We recommend introducing interactive tools that explain each disclosure step with contextual examples and consistent definitions. 
Clearer terminology and examples would reduce errors and improve confidence.

%We recommend the development of interactive tools that not only detect data collection but also explain the \emph{rationale} for each required disclosure. 
%Step-by-step guidance, contextual examples, and decision aids could significantly reduce developer confusion and errors. 
%Additionally, publishing consistent and transparent definitions would help developers navigate the form with greater confidence and ensure more accurate reporting. 
%AutoPRICE implements this by progressively revealing justifications for each data type included in the pre-filled DSS form (\textcircled{3} in \Cref{fig:filleddss}). 
%This feature was well-received by participants in our user study (\Cref{US}), suggesting potential for broader adoption. 

\textbf{Existing Tools and their Limitations.} Although tools to support DSS completion already exist~\cite{googlechecks,matcha,privadoai}, adoption remains limited. 
Only 18 of our survey participants (43.9\%) reported using such tools. 
Of these, six still categorized collected data manually, and five omitted categorization altogether. 
Among the 13 who used Google's Checks framework~\cite{googlechecks}, five found DSS completion easy, while five described it as challenging. 
P32, who struggled the most despite using Checks, said, \directquote{Understanding issues about the rejection and solving the problem [is the most challenging]}. 
P39, a Privado~\cite{privadoai} user, said, \directquote{The manual work required is a bit challenging. I'm always a bit skeptical as to whether I've accurately captured the data safety details in Play Store.} 
P40, who used Checks alongside an in-house tool, said, \directquote{Accurately identifying and categorizing all types of user data collected by the app (e.g., location, personal information, device data) can be challenging, especially in apps with third-party SDKs.} 
A user in one online discussion on Reddit admitted to having tried Privado in their second attempt to make sure their DSS reported all data types their app collects, but still got an ``undeclared data collection'' rejection message from Google.  

Survey responses showed developers' reluctance to use existing tools. 
For instance, P40 said, \directquote{Many developers lack tools to automatically track and audit data usage, making it challenging to identify all collected or shared data.} 
P20 said, \directquote{I use ChatGPT for that (completing the DSS form) since many things are known only to jurists, who cost a lot, especially for private developers.}  
%I try to avoid collecting any data. 
%It makes the user experience worse but affordable.}\todo{This could be a finding - some developers avoid collecting data even compromising the user experience to keep filling DDS form affordable.}

%Despite these challenges, only 18 of our survey participants (43.9\%) reported using third-party or in-house tools, with 13 participants (31.71\%) having used Google’s Checks framework~\cite{googlechecks}.  
Community adoption appears similarly low: only 15 discussions (8.72\%) referenced third-party tools such as Privado~\cite{privadoai}, with three posts authored by Privado’s creators themselves—suggesting limited community uptake of such solutions. 

These findings suggest that current DSS-support tools may fall short in terms of accuracy, usability, or accessibility. 
%They may not always help developers navigate rejection errors from Google Play Store. 
Future work should systematically evaluate these tools to understand why they are underutilized and how they can be improved to better support app developers. 
%P20 admitted to using ChatGPT and added, \directquote{There many things which are known only to jurists, which are costly, especially for private developers. I try to avoid collecting any data. It makes the user experience worse, but affordable.} 

Improving the DSS ecosystem therefore requires a multifaceted approach: clearer, standardized guidelines from Google; better alignment between SDK documentation and actual behaviors; and practical, developer-centered tools that make disclosure both accurate and accessible. 
Ultimately, bridging these gaps would not only improve the reliability of DSS disclosures but also strengthen user trust in the privacy information presented on the Play Store.

%% file: Sections/RelatedWork.tex
\section{Related Work}
\label{relatedwork}

We now discuss various aspects of existing related research. % on the topic.

\textbf{DSS Accuracy.} Multiple studies have examined the accuracy of DSS submissions~\cite{datalabels,khedkar2024androidappdevelopersaccurately,mozilla,10190677}.  
A Mozilla study~\cite{mozilla} evaluated popular Android apps and found discrepancies between their DSS and privacy policies. 
Khandelwal et al.~\cite{datalabels} explored how developers engage with the DSS by analyzing snapshots of submitted forms and contacting over 3,500 developers. 
More recently, Girish et al.~\cite{girish2025signaldataempiricalprivacy} compared the runtime behavior of beacon-enabled apps with their declared safety labels, revealing that many apps collect device identifiers without proper disclosure. 
While these studies collectively highlight persistent challenges in the disclosure process, they do not specifically investigate \emph{how} developers report data collection—a stage that has proven particularly error-prone~\cite{khedkar2024androidappdevelopersaccurately,girish2025signaldataempiricalprivacy} (\textbf{RQ1} of our study). 
Our work extends this line of inquiry by not only exploring the challenges developers face in reporting data collection (\textbf{RQ3}), but also analyzing their confidence levels (\textbf{RQ2}). 

\textbf{Analyzing Online Discussions.} Prior research~\cite{SOreddit, SOreddit1, developercommunities, survey3} has examined online developer communities such as Stack Overflow and Reddit to study various aspects of software development practices. 
To the best of our knowledge, our work is the first to use discussions from these online communities to explore the specific challenges Android developers face when completing the DSS form. 

\textbf{DSS-Support Tools.} Prior studies evaluating DSS accuracy~\cite{datalabels,mozilla,khedkar2024androidappdevelopersaccurately} were conducted at a time when DSS-assistive tools were either unavailable or too immature to study in practice, leaving open questions about their potential impact on developers’ disclosure practices.   
Since then, Google has introduced Checks~\cite{googlechecks}, a paid service that assists app developers in completing the DSS, alongside open-source alternatives such as Matcha~\cite{matcha} and Privado.ai~\cite{privadoai}. 
Matcha analyzes app code and suggests appropriate DSS labels, relying on developer annotations to finalize the disclosures. 
Privado is a command-line tool that accepts open-source projects as input, and then generates their DSS form. 
Despite these advancements, little is known about how widely such tools are adopted, whether they effectively address challenges in reporting data collection, or what limitations remain. 
By examining the adoption of these tools (\textbf{RQ1}), and analyzing the broader challenges developers face in data reporting (\textbf{RQ3}), our study provides timely insights into a problem space that has evolved considerably since earlier work.

%% file: Sections/Limitations.tex
\section{Limitations and Threats to Validity}
\label{limitations}

%\stilltodo{TODO: Details T2V for empirical study?}

%The study presented in this paper has limitations that could potentially jeopardize the validity of the results.
%\textbf{Empirical Study.} 
In this section we discuss the limitations of the study performed in this paper and consider implications to the validity of the results.

Our analysis centers exclusively on Google’s DSS, and hence its results cannot be generalized to Apple’s labels without further research. 
Conversely, findings from prior work on Apple’s labels~\cite{survey2} are not directly transferable to our context. 
Our results can motivate and guide researchers to replicate the study for Apple Store. 

The developer survey is susceptible to self-selection and recall biases, which may influence the representativeness and accuracy of responses. 
The sample size (41 developers) also limits generalizability. 
To mitigate these risks, we conducted multiple rounds of recruitment and directly contacted 196 developers to ensure diversity across app domains and organization types. 
Nonetheless, participation was constrained by the nature of the task—several developers declined due to reliance on compliance or app publishing teams for completing the DSS form. 
To strengthen external validity and compensate for the modest survey size, we supplemented the survey findings with data from online developer discussions. 

Our analysis of developer feedback in online communities focuses on developer discussions from Stack Overflow, Reddit, Discord, GitHub, and Hacker News, which may not reflect perspectives from all communities. 
However, prior research~\cite{reddit1,reddit2, SOreddit, SOreddit1} supports online community's  value for qualitative insights. 
To broaden our coverage, we also examined other popular developer forums (Dzone~\cite{dzone}, Sitepoint~\cite{sitepoint}, Code Project~\cite{codeproject}, and XDA forums~\cite{xdaforums}) but found no relevant posts. 
Since the DSS was introduced in 2022~\cite{dssintroblog}, the volume of related discussions remains limited. 
To address this, we ensured that our dataset included sufficient engagement (642 unique developers), and we complemented it with survey data to enable triangulation~\cite{triangulation}. 
Our data collection concluded in September 2025 and includes the most recent discussions available at the time of analysis. 
Nonetheless, as DSS adoption continues to evolve, future research should examine emerging discussions to capture potential changes in developer experiences.

To ensure that our analysis accurately represented developers’ experiences with the DSS, we employed an iterative open coding process~\cite{opencoding} following thematic analysis principles~\cite{thematic}. The first author performed inductive coding to derive emergent themes. 
To improve coding consistency and reliability, the second author independently reviewed the annotations, and disagreements were resolved through discussion, ensuring coder alignment. 
This systematic refinement process, along with the public availability of the final codebook (\Cref{codebook}), enhances the transparency and construct validity of our findings by ensuring that the themes faithfully reflect the underlying data. 

Despite these limitations, the convergence between survey findings and online discussion analysis provides confidence in the validity of our results, suggesting that the combined evidence base is sufficient to draw meaningful insights. 

%% file: Sections/Conclusion.tex
\section{Conclusion}
\label{conclusion}

In this paper, we conducted an empirical study to understand Android developers' experience with the DSS form. 
Through a survey of 41 Android app developers, we found that developers often manually classify the privacy-related data their apps collect into the DSS data categories defined by Google—or, in some cases, omit classification entirely—and rely heavily on existing online resources when completing the DSS forms (\textbf{RQ1)}. 
While developers generally feel confident in identifying the data their apps collect, their confidence drops significantly when translating this knowledge into DSS-compliant disclosures (\textbf{RQ2}).

To broaden our understanding of these challenges, we analyzed 172 online developer discussions, revealing recurring issues such as issues in identifying privacy-relevant data to complete the form, limited understanding of the form, and concerns about app rejection due to discrepancies with Google’s privacy requirements (\textbf{RQ3}). 

Overall, our findings highlight the importance of more usable tooling and clear guidance to support Android developers in meeting privacy-aware reporting requirements. 